\documentclass[amsmath, amsthm, amssymb, aps, prb, superscriptaddress, twocolumn, nofootinbib, 10pt]{revtex4-1}

\makeatletter
\def\@seccntformat#1{\csname the#1\endcsname.\quad}
\renewcommand\section{\@startsection {section}{1}{0pt}%
  {-3.5ex \@plus -1ex \@minus -.2ex}%
  {2.3ex \@plus.2ex}%
  {\normalfont\normalsize\bfseries\raggedright}}
\renewcommand\subsection{\@startsection {subsection}{2}{0pt}%
  {-3.25ex\@plus -1ex \@minus -.2ex}%
  {1.5ex \@plus .2ex}%
  {\normalfont\normalsize\bfseries\raggedright}}
\makeatother



\newcommand{\inlinecite}[1]{\citenum{#1}}

\usepackage[colorlinks=true,linkcolor=blue,citecolor=blue,urlcolor=blue]{hyperref}
\usepackage{amsmath,amssymb,amsfonts}
\usepackage{graphicx}
\usepackage{cleveref}
\crefname{figure}{Fig.}{Figs.}

\usepackage{siunitx}
\usepackage[resetlabels,labeled]{multibib}
\usepackage{etoolbox}
\usepackage{float}
\usepackage{geometry}
\usepackage{natbib}
\usepackage{mathrsfs}
\usepackage{textcomp}
\usepackage{lipsum}
\usepackage{xcolor}
\usepackage{upgreek}
\usepackage{gensymb}

\begin{document}

\title{\LARGE{A scalable quadratic nonlinear silicon photonics platform with printable entangled photon-pair sources}}

\author{Tom~Vandekerckhove$^{1,2,3,*}$, Jasper~De~Witte$^{1,2}$, Lisa~De~Jaeger$^{1,2}$, Ewoud~Vissers$^{1,2}$, Sofie~Janssen$^{2}$, Peter~Verheyen$^{2}$, Neha~Singh$^{2}$, Dieter~Bode$^{2}$, Martin~Davi$^{2}$, Filippo~Ferraro$^{2}$, Philippe~Absil$^{2}$, Sadhishkumar~Balakrishnan$^{2}$, Joris~Van~Campenhout$^{2}$, Dries~Van~Thourhout$^{1,2}$, G\"unther~Roelkens$^{1,2}$, St\'ephane~Clemmen$^{1,2,3}$ and Bart~Kuyken$^{1,2,*}$\\  
\vspace{+0.1 in}
\textit{\small{
$^1$Photonics Research Group, INTEC, Ghent University - imec, 9052 Ghent, Belgium\\
$^2$imec, Kapeldreef 75, 3001 Leuven, Belgium. \\
$^3$OPERA-Photonique CP 194/5, Université Libre de Bruxelles, 1050 Brussels, Belgium\\
{\small $*$Tom.Vandekerckhove@ugent.be, Bart.Kuyken@ugent.be}}}}

\begin{abstract} 
\textbf{
The integration of second-order optical nonlinearities into scalable photonic platforms remains a key challenge due to their large sensitivity to fabrication variations. Here, we present a scalable quadratic nonlinear platform that harnesses the maturity and scalability of existing CMOS processes by heterogeneously integrating periodically poled lithium niobate (PPLN) onto a silicon photonics platform. A generic PPLN design enables frequency conversion on two distinct waveguide geometries with efficiencies comparable to LNOI rib waveguides. We achieve reproducible phase-matching across the full radius of a commercial 200~mm silicon photonics wafer, leveraging superior CMOS fabrication tolerances. Furthermore, we introduce a tuning mechanism for both blue- and red-shifting of the operating wavelength, fully compensating fabrication-induced offsets. This enables deterministic phase-matching over an entire wafer and yields a strategy for wafer-scale phase-matched quadratic nonlinearities. Finally, we realize printable photon-pair sources via spontaneous parametric down-conversion, highlighting the platform's potential for large-scale quantum optical circuits. These results pave the way for wafer-scale integration of second-order optical nonlinearities in large photonic systems.
}
\vspace{-0.5cm}
\begin{figure}[b!]
\centering
\includegraphics[width=\linewidth]{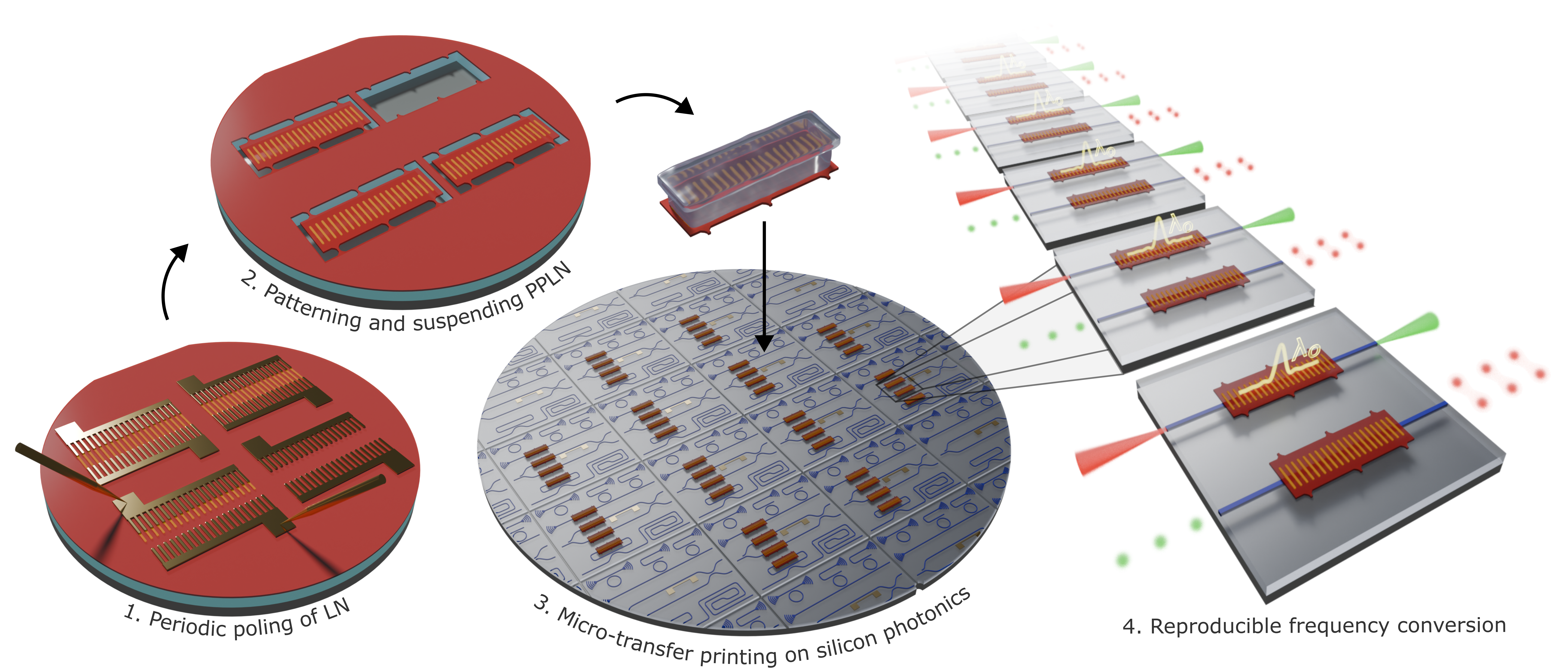}
\caption{\textbf{A scalable quadratic nonlinear silicon photonics platform.} \textbf{1} Thin-film lithium niobate is periodically poled through electric field poling. \textbf{2} PPLN films are patterned and suspended. \textbf{3} PPLN films are micro-transfer printed on exposed silicon nitride waveguides. \textbf{4} Reproducible frequency conversion and photon-pair generation are obtained.}
\label{fig_platform sketch}
\vspace{0.5cm}
\end{figure}
\end{abstract}

\maketitle


\begin{figure*}[t]
\centering
\includegraphics[width=\linewidth]{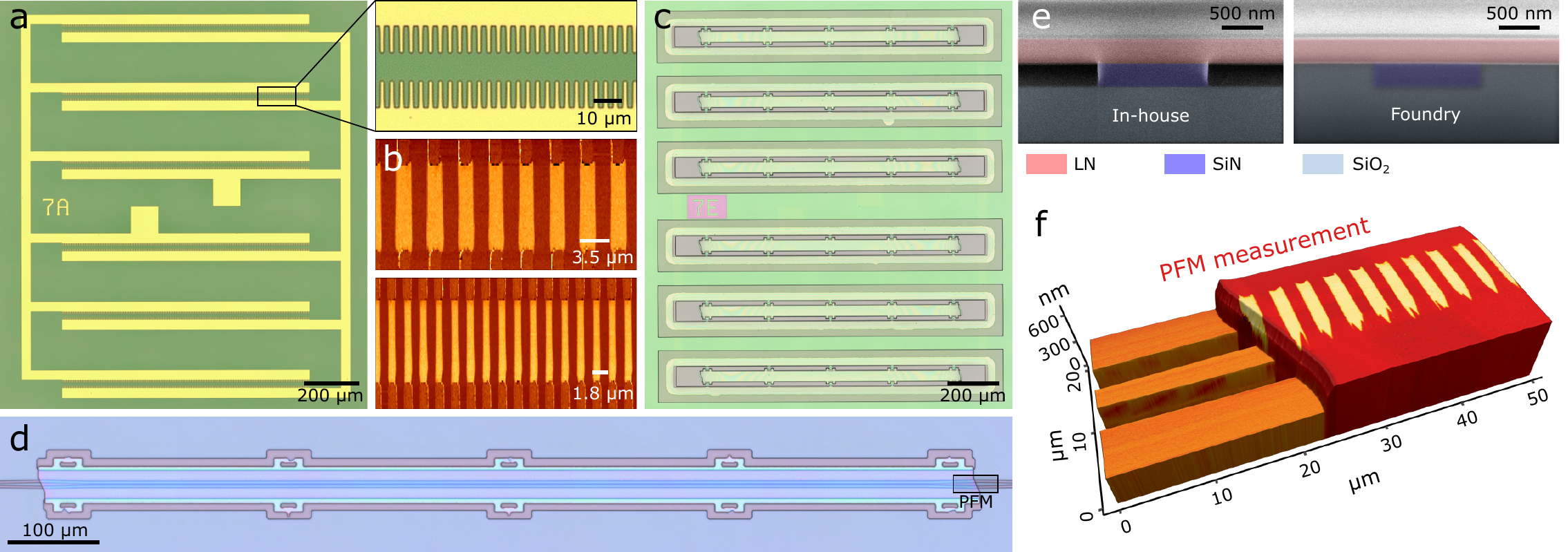} 
\caption{\textbf{Platform process flow.} \textbf{a} Optical microscope image of deposited electrodes for electric field poling of 6 LN films. \textbf{b} PFM images visualizing the straight poled domains for a 3.5 $\upmu$m (upper) and 1.8 $\upmu$m (lower) poling period. \textbf{c} Optical microscope image of suspended PPLN thin films, ready to be picked. \textbf{d} Optical microscope image of micro-transfer printed PPLN film onto a silicon nitride waveguide, fabricated in-house. \textbf{e} False-color SEM images of PPLN/SiN waveguide cross sections on the in-house (left) and foundry (right) platform (with Pt deposited on top for a cleaner ion-beam cut). \textbf{f} Visualization of an AFM measurement with a PFM color overlay, measured at the transition from a silicon nitride waveguide (indicated on Fig.\ref{fig_process flow}d). This demonstrates that the poled domains are positioned above the nitride waveguide.}
\label{fig_process flow}
\end{figure*}

Over the past decade, the efficiency of optical frequency conversion processes has significantly improved, driven by the emergence of the lithium niobate on insulator (LNOI) platform\cite{boes2023lithium, zhang2017monolithic}. The introduction of etched LN waveguides with quadratic nonlinearities has enabled sub-wavelength mode confinement, leading to nonlinear conversion efficiencies more than 20 times higher than those of previous in-diffused waveguides \cite{wang2018ultrahigh, rao2019actively, park2022high}. Moreover, new nonlinear regimes are unlocked due to the strong geometric dispersion associated with the high index contrast. By leveraging the ability to periodically pole the lithium niobate, quasi-phase-matching can be achieved while simultaneously engineering higher-order dispersion through waveguide design. This advancement has led to several demonstrations, including ultrabroadband frequency conversion \cite{jankowski2020ultrabroadband, zhang2022ultrabroadband}, large-gain optical parametric amplification \cite{jankowski2022quasi} and spectrally-separable photon-pair generation \cite{xin2022spectrally}. The nonlinear response can be further enhanced by incorporating the periodically poled lithium niobate (PPLN) into a resonator, yielding normalized conversion efficiencies up to 5,000,000~\%/W \cite{lu2020toward} and low-threshold optical parametric oscillators \cite{lu2021ultralow,mckenna2022ultra}.

Unfortunately, the same geometric dispersion leads to significant fabrication sensitivity of the phase-matching condition \cite{kuo2021noncritical,zhao2023unveiling}. Non-uniformity along the waveguide limits the achieved conversion efficiencies while fabrication deviations from the intended geometry cause a substantial shift in the operating wavelength, preventing deterministic phase-matching. Solutions have been proposed based on in-depth characterization of the waveguide geometry through advanced metrology, followed by locally adapting the poling period\cite{chen2024adapted,xin2024wavelength} or waveguide dimensions \cite{he2024efficient} in subsequent fabrication steps. However, this severely limits the scalability to larger volumes, as every wafer requires a custom lithography mask, which is time-consuming and expensive.

In contrast, the silicon photonics platform offers much better fabrication tolerances and is inherently scalable, benefiting from well-established CMOS manufacturing techniques\cite{shekhar2024roadmapping}. Especially, the silicon nitride platform has gained popularity for nonlinear and quantum applications due to its large transparency window, low losses and power-handling capabilities\cite{blumenthal2018silicon, buzaverov2024silicon}. Nevertheless, both silicon and silicon nitride waveguides lack a $\chi^{(2)}$ component due to their inversion symmetry, inhibiting efficient frequency conversion. Quadratic nonlinearities can be induced through symmetry-breaking methods\cite{billat2017large, timurdogan2017electric}, but the conversion efficiencies are typically much weaker than the LNOI platform.

Heterogeneous integration of lithium niobate on a silicon photonics platform is able to combine the high nonlinear efficiencies of the LNOI platform with the fabrication tolerances and scalability of silicon photonics. While several demonstrations have successfully integrated thin-film lithium niobate\cite{vanackere2023heterogeneous, churaev2023heterogeneously, chang2017heterogeneous, ruan2023high}, effective use of the $\chi^{(2)}$ component through a phase-matched process remains elusive. 

In this work, we establish a nonlinear silicon photonics platform by micro-transfer printing PPLN onto silicon photonics (see Fig.\ref{fig_platform sketch}). This method enables back-end integration of thin-film lithium niobate, preventing lithium contamination\cite{wandesleben2024influences} during CMOS processing while preserving the scalability of the silicon photonics platform\cite{roelkens2024present}. We demonstrate quasi-phase-matched frequency conversion with comparable efficiencies to LNOI rib waveguides on two distinct silicon photonics platforms with a generic PPLN design. Furthermore, introducing a dual-waveguide design allows monitoring of frequency converters in closed optical circuits, marking an important step towards larger and more complex nonlinear systems. Next, reproducibility is shown over a complete radius of a 200~mm silicon photonics foundry wafer, showcasing full control by blue- and red-shifting the operating wavelength. Based on these results, we propose a strategy for a scalable quadratic nonlinear silicon photonics platform with reproducible and deterministic phase-matching. Additionally, we establish printable entangled photon-pair sources by demonstrating spontaneous parametric down-conversion (SPDC) in the proposed platform.

Our approach paves the way for scalable quadratic nonlinearities in large-scale nonlinear photonic circuits, from arrays of interfering parametric photon-pair sources to the scalable fabrication of low-threshold optical parametric oscillators. 

\begin{figure*}[t]
\centering
\includegraphics[width=\linewidth]{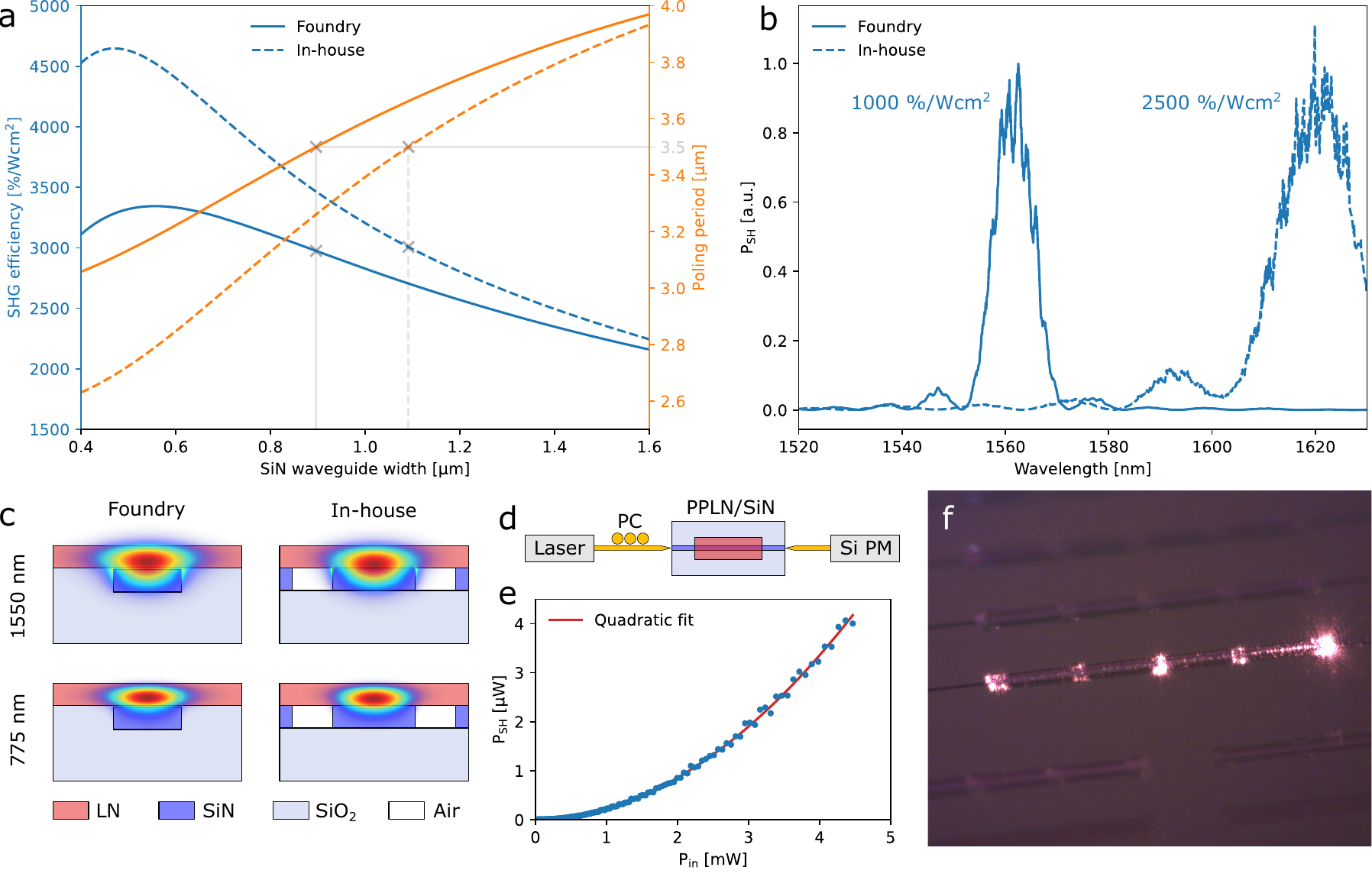}
\caption{\textbf{Quasi-phase-matched frequency conversion}. \textbf{a} Numerical simulation of the SHG efficiency and corresponding poling period as function of the silicon nitride width, for a pump at 1550 nm. The chosen 3.5~$\upmu$m period is indicated, giving SiN widths of 900~nm (foundry) and 1100~nm (in-house). \textbf{b} Measured SHG frequency conversion on foundry and in-house platform. \textbf{c} Mode profiles of the fundamental TE00 and second harmonic TE00 in the hybrid waveguides. \textbf{d} Sketch of the fiber-based measurement setup with polarization controller (PC), silicon power meter (PM) and lensed fibers. \textbf{e} Measured quadratic power dependence of the frequency conversion. \textbf{f} Photograph of the generated second harmonic in a PPLN/SiN waveguide.}
\label{fig_SHGcharacterisation}
\end{figure*}

\section*{Results}
\subsection*{Fabrication of the nonlinear waveguide}

The process flow of our nonlinear silicon photonics platform begins with the periodic poling of X-cut thin-film lithium niobate. Starting from an LNOI wafer (300~nm LN thickness), electrodes are deposited for electric field poling with a period matching the designed phase-matching process (Fig.\ref{fig_process flow}a). Similar to previous works \cite{stanicki2020surface, zhao2020poling}, a custom electric waveform is applied to control domain growth vertically while keeping horizontal growth limited. This ensures the formation of straight domains across various poling periods (3.5~$\upmu$m and 1.8~$\upmu$m shown in Fig.\ref{fig_process flow}b), which are characterized through piezoresponse force microscopy (PFM). Due to the pick-and-place approach, positioning of the poling electrodes is independent of the final photonic circuit layout. This allows us to arrange our nonlinear devices in a dense array or matrix configuration, where the electrodes are connected in parallel without crossing any future waveguides. A single waveform then poles a full set of devices simultaneously, making it possible to efficiently scale the poling process to larger volumes.

After poling the lithium niobate, the PPLN is patterned into rectangular films of 40~$\upmu$m~$\times$~1~mm and suspended by etching the buried oxide layer with hydrofluoric acid (Fig.\ref{fig_process flow}c). Only a few anchoring points keep the devices connected to the substrate. A polymer PDMS stamp is then used to pick up the films with a commercial micro-transfer printing tool, breaking these connections and fully detaching them. Next, they are printed onto exposed 300~nm thick silicon nitride waveguides, forming a hybrid PPLN/SiN geometry (Fig.\ref{fig_process flow}d). An in-depth description of the transfer printing flow is given by Ref.~\inlinecite{vandekerckhove2023reliable}. Two silicon nitride platforms are used: an in-house prototyping platform based on electron beam lithography and a 200~mm silicon photonics foundry platform\cite{ferraro2023imec}. Cross sections of the fabricated hybrid waveguides are shown in Fig.\ref{fig_process flow}e, where the in-house platform features 3~$\upmu$m-wide air trenches, while the foundry platform is fully planarized. A simultaneous atomic force microscopy (AFM) and PFM measurement at the transition to the hybrid waveguide confirms that the poled domains are aligned above the silicon nitride waveguide (Fig.\ref{fig_process flow}f). The fabricated devices incorporate angled waveguide transitions at 25$\degree$ to avoid parasitic reflections, which dominate the insertion losses of the component (see supplementary Fig.\ref{suppfig_angledtransition}). The outcome is a quasi-phase-matched nonlinear waveguide on a silicon photonics platform.

\begin{figure*}[t]
\centering
\includegraphics[width=\linewidth]{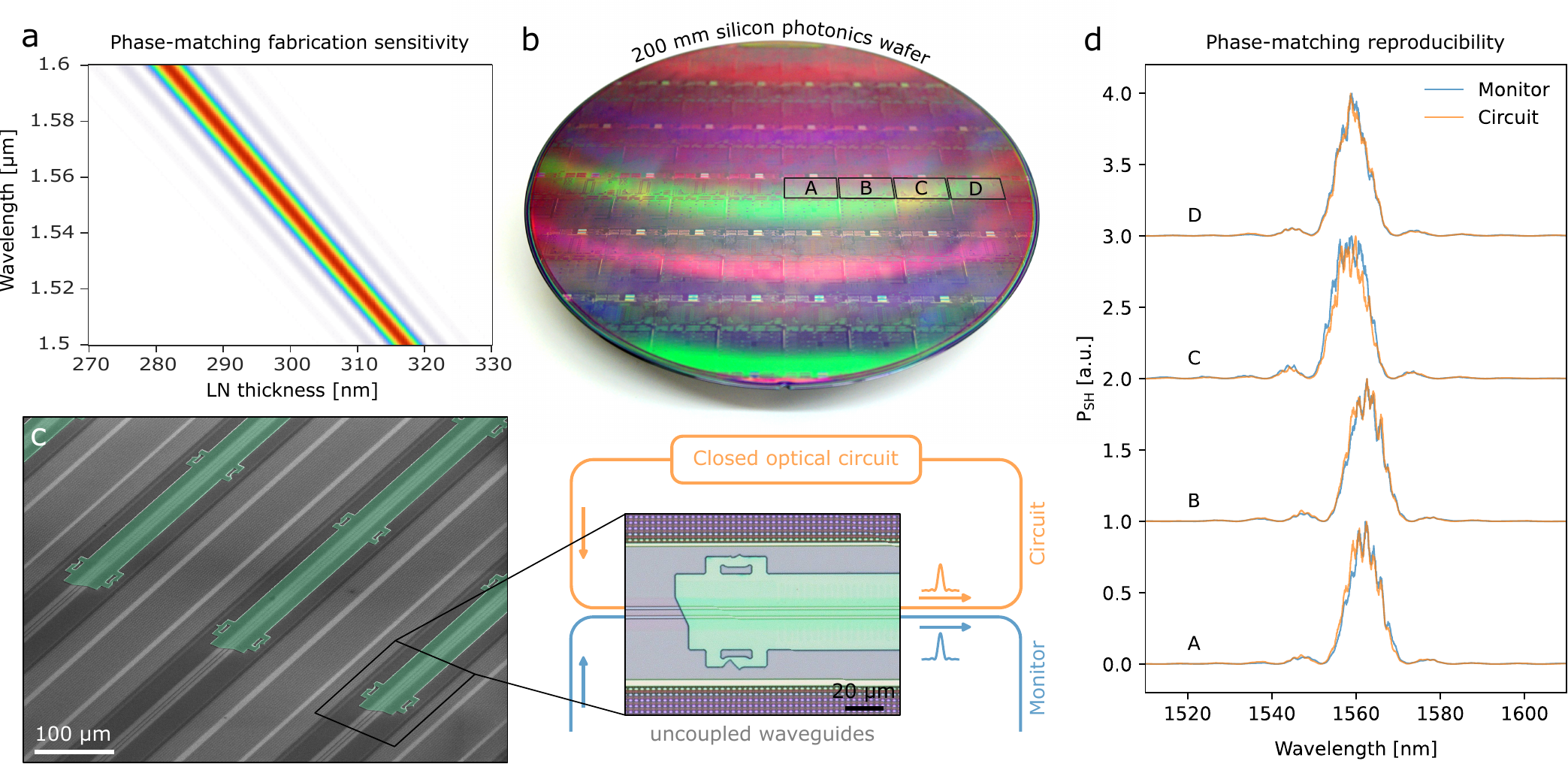}
\caption{\textbf{Reproducible phase-matching on a 200 mm silicon photonics wafer.} \textbf{a} Simulated SHG phase-matching sensitivity to the LN thickness. \textbf{b} Photograph of the 200 mm silicon photonics foundry wafer. \textbf{c} False-color SEM image of printed PPLN films on the foundry platform, with an optical microscope image of the dual-waveguide design relying on two uncoupled (circuit and monitor) waveguides sharing the same PPLN film. \textbf{d} Measured phase-matching reproducibility with identical phase-matching between monitor and circuit waveguides, and nearly-identical phase-matching over the full 200 mm wafer radius (indicated on Fig.\ref{fig_foundrySiPh}b).}
\label{fig_foundrySiPh}
\end{figure*}

\begin{figure*}[t]
\centering
\includegraphics[width=\linewidth]{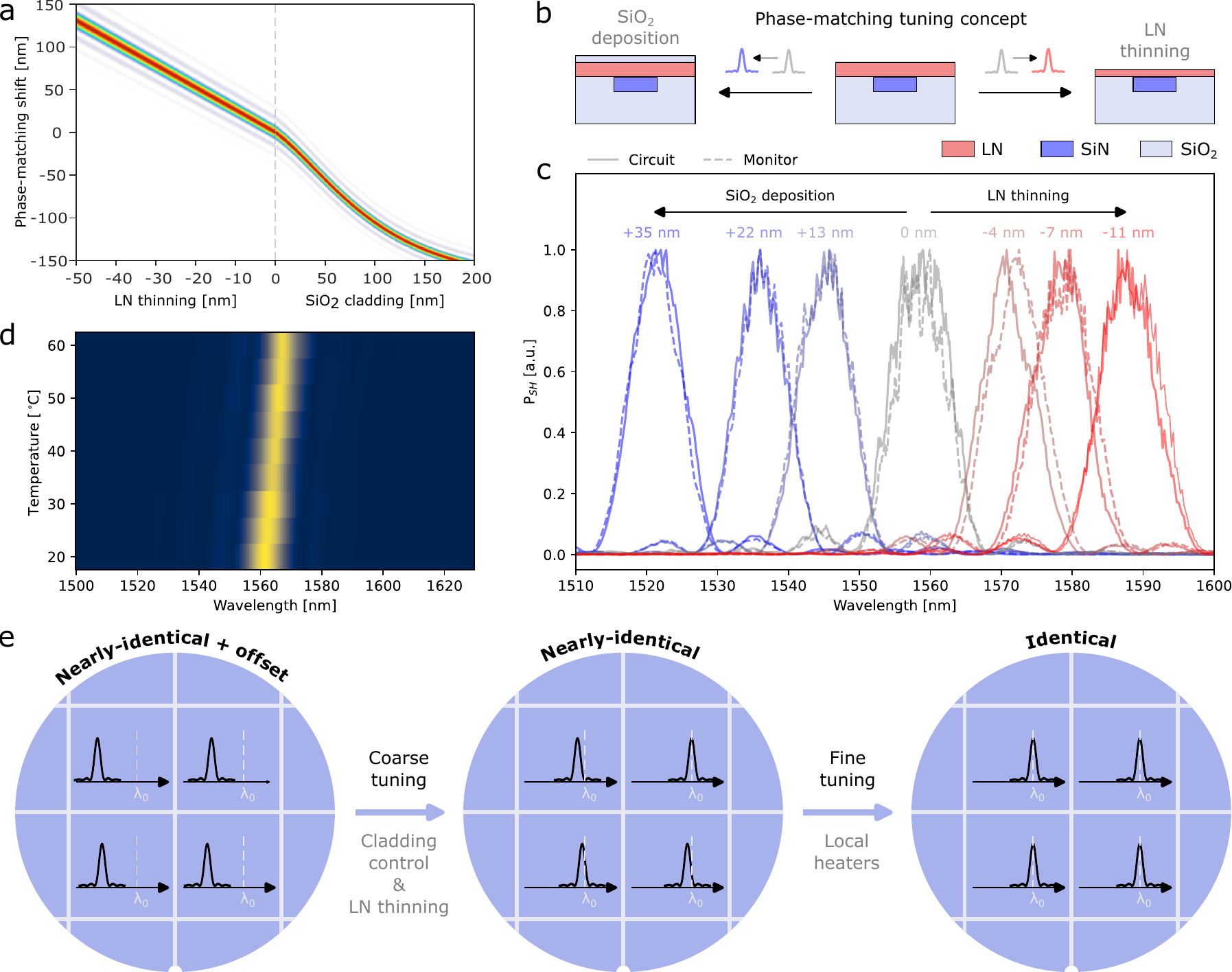}
\caption{\textbf{Phase-matching tunability for wafer-scale nonlinearities.} \textbf{a} Simulated phase-matching tuning through SiO$_2$ cladding or LN thinning. \textbf{b} Phase-matching tuning concept through waveguide modification. \textbf{c} Measured simultaneous tuning of circuit and monitor waveguide through cladding control (blue-shift) and LN thinning (red-shift). \textbf{d} Measured temperature dependence of the phase-matching with a 0.18~nm/K efficiency. \textbf{e} Strategy to obtain identical phase-matching over a full silicon photonics wafer though coarse (cladding control \& LN thinning) and fine (local heaters) tuning.}
\label{fig_PMtuning}
\end{figure*}

\subsection*{Efficient frequency conversion on distinct silicon photonics platforms}

Our nonlinear silicon photonics platform decouples the poling process from the photonic circuit design, offering increased flexibility. By adapting the nitride waveguide width, the same PPLN film can be used to obtain phase-matching at a given wavelength for different layer stacks (see supplementary Fig.\ref{suppfig_genericPPLN}). This approach enhances scalability, as an entire wafer of identical PPLN films can be leveraged to accommodate phase-matching requirements across multiple silicon photonics platforms, each operating at several wavelengths. 

We demonstrate this versatility by employing the same generic PPLN design for our in-house and foundry platform. A LN thickness of 300 nm and poling period of 3.5~$\upmu$m are taken to enable phase-matching on both platforms for waveguide widths around 900~nm and 1100~nm, respectively (Fig.\ref{fig_SHGcharacterisation}a). These amount to simulated nonlinear efficiencies of about 3000~$\%$/Wcm$^2$ for the type-0 second harmonic generation (SHG) process at 1550~nm. The involved TE00 modes are shown in Fig.\ref{fig_SHGcharacterisation}c. After fabrication, we achieve quasi-phase-matched SHG on both platforms, at 1620 nm (in-house) and at 1560 nm (foundry) (Fig.\ref{fig_SHGcharacterisation}b). A deviation in the LN thickness from the intended 300~nm causes a notable shift from the 1550~nm target wavelength of the former, highlighting the need for a strategy to achieve deterministic phase-matching, as discussed further. Nevertheless, quasi-phase-matching is observed with conversion efficiencies of 2500~$\%$/Wcm$^2$ (in-house) and 1000~$\%$/Wcm$^2$ (foundry), comparable to LNOI rib waveguides \cite{wang2018ultrahigh}. 3-dB bandwidths of 15~nm and 8~nm are measured, respectively, related to the higher-order dispersion of the waveguide geometries. Furthermore, a measured quadratic power dependence, along with direct imaging of the second harmonic build-up in the PPLN/SiN waveguide, confirms the SHG process (Fig.\ref{fig_SHGcharacterisation}e and \ref{fig_SHGcharacterisation}f). These results demonstrate the successful heterogeneous integration of quasi-phase-matched quadratic nonlinearities on distinct silicon photonics platforms.

\begin{figure*}[t]
\centering
\includegraphics[width=\linewidth]{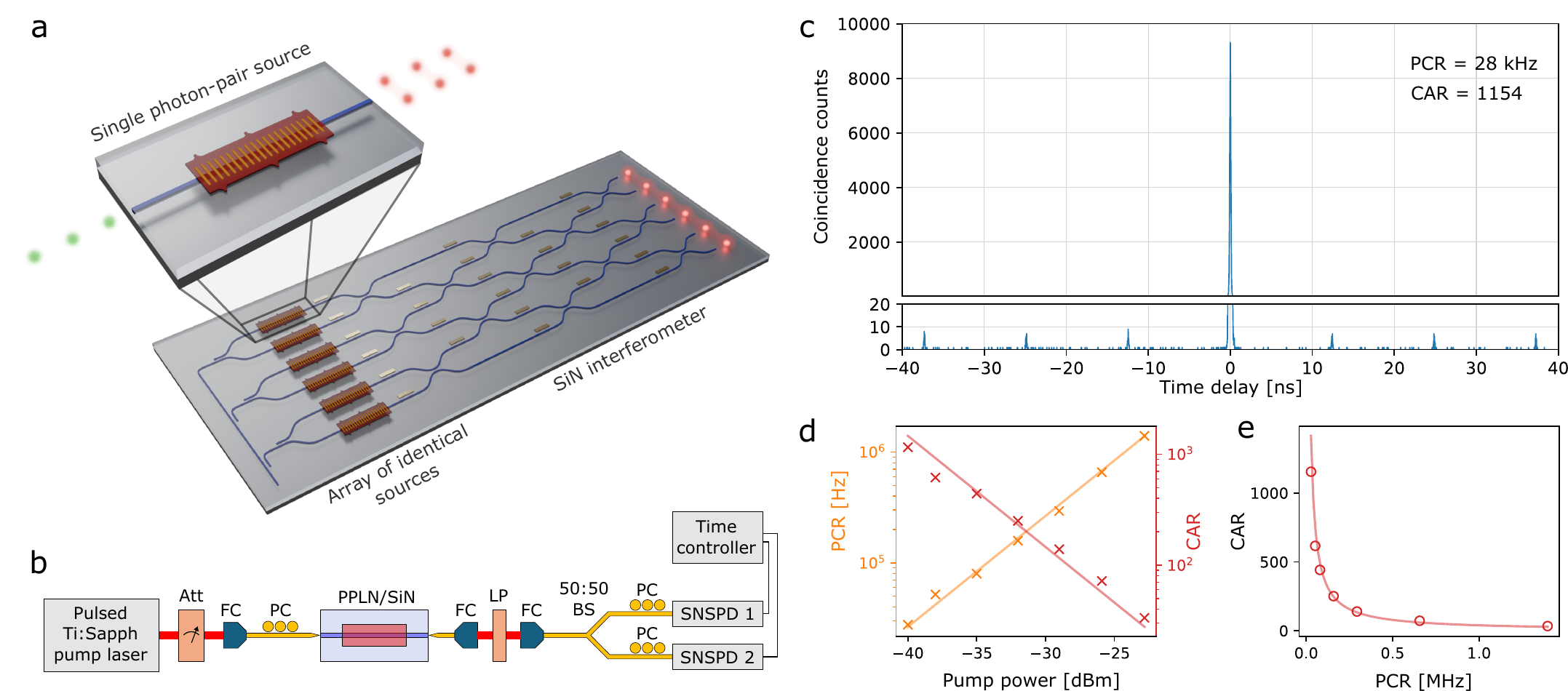}
\caption{\textbf{Printable entangled photon-pair sources.} \textbf{a} Conceptual sketch of a large photonic quantum system with an array of identical printed photon-pair sources in front of a large SiN interferometer. \textbf{b} Hanbury-Brown-Twiss measurement setup with variable attenuator, fiber collimators (FC), lensed fibers, longpass filter (LP), beam splitter (BS), polarization controller (PC) and superconducting nanowire single-photon detectors (SNSPD). \textbf{c} Measured g$^{(2)}$ correlation histogram for a coincidence-to-accidental ratio (CAR) of 1154. \textbf{d} On-chip pair coincidence rate (PCR) and measured CAR as function of on-chip average pump power. \textbf{e} Measured CAR as function of on-chip PCR.}
\label{fig_SPDC}
\end{figure*}


\subsection*{Chip-to-chip reproducibility on a 200 mm silicon photonics wafer}

Current LNOI rib waveguides struggle with phase-matching reproducibility due to their high sensitivity to fabrication variations, which stem from two main sources. First, the chemical mechanical polishing during LNOI manufacturing leads to typical LN film thickness variations of 10-20~nm across the wafer\cite{xin2024wavelength}. Second, the subsequent shallow etch used to define the rib waveguides suffers from a challenging process control, resulting in additional etch depth variations and non-uniformity on the order of 20~nm\cite{xin2024wavelength}. Consequently, large deviations are observed from the intended phase-matching wavelength.

Our hybrid approach seeks to improve reproducibility by relying on silicon nitride strip waveguides, which benefit from superior fabrication tolerances due to their mature wafer-scale CMOS processes on 200~mm and 300~mm wafers\cite{siew2021review, zhang2024300}. This eliminates variations associated to the shallow etch, leaving only original LN thickness deviations as dominant factor. Additionally, as the micro-transfer printing method enables dense packing of PPLN films on the LNOI source wafer (Fig.\ref{fig_process flow}c), neighboring films from the same local area can be used to populate the silicon photonics wafer. This minimizes differences in thickness between devices and further improves reproducibility over the wafer. 

Generally, a discrepancy from the intended film thickness will remain, introducing a consistent offset in phase-matching over the full wafer. For every nanometer difference in thickness, a shift of 2.8~nm (5.0~nm) in phase-matching wavelength is expected for the foundry (in-house) nitride platform (Fig.\ref{fig_foundrySiPh}a and supplementary Fig.\ref{suppfig_nonidealfabrication}), signaling the need for a tuning mechanism to achieve deterministic phase-matching, as discussed further. Direct access to the nonlinear waveguide is then typically desired for optical characterization of the offset, which imposes constraints on the design of more complex nonlinear circuits. We circumvent this issue by adopting a dual-waveguide approach (Fig.\ref{fig_foundrySiPh}c). Two parallel but uncoupled waveguides are covered with the same PPLN film, ensuring they have identical domain structures and waveguide geometry along their entire length. As a result, both waveguides are expected to exhibit the same phase-matching behavior, enabling one to be integrated into a closed optical circuit while the other functions as a dedicated monitor.

To assess the reproducibility of our platform, we fabricate nonlinear devices over the full radius of a 200 mm silicon photonics wafer, designed for second harmonic generation at 1550~nm. PPLN films are picked from a single LNOI source and printed onto the four chips that constitute the wafer radius, as indicated in Fig.\ref{fig_foundrySiPh}b. The corresponding optical characterization for each chip is shown in Fig.\ref{fig_foundrySiPh}d. A perfect overlap is observed between the phase-matching curves of each monitor and circuit waveguide. This validates the dual-waveguide approach, where a monitor can reliably probe the phase-matching in a closed nonlinear circuit. Comparing the different chips across the wafer radius reveals nearly-identical phase-matching for the same waveguide designs. The minor remaining shifts can be fully attributed to the original non-uniformity of the LNOI source, which would amount to 1-2~nm according to simulations. Furthermore, the micro-transfer printing method introduces no detectable variations between different printing steps. Although tested along a single radius, the measured reproducibility strongly suggests that nearly-identical phase-matching can be achieved over the entire 200 mm wafer. 

\subsection*{Tunability for wafer-scale deterministic phase-matching}

While consistent across the wafer, there remains an offset relative to the designed wavelength, indicating the need for a tuning mechanism to reach deterministic phase-matching. Thermo-optic tuning is a common approach to shift the phase-matching, and characterization of our platform reveals a tuning efficiency of 0.18~nm/K (Fig.\ref{fig_PMtuning}d). Unfortunately, fully neutralizing offsets that can exceed 50 nm would require impractically high temperatures, making this method unfeasible by itself. An alternative tuning approach is therefore necessary.

We propose a single-step process for simultaneous phase-matching tuning over a full wafer. Although the high fabrication sensitivity introduces a consistent offset, it also implies that small controlled modifications to the waveguide geometry can create significant wavelength shifts. The hybrid waveguide geometry is particularly well-suited for this, as it features a single accessible flat interface while the nitride waveguide remains protected. Effectively altering the LN thickness then allows for significant phase-matching shifts (Fig.\ref{fig_PMtuning}b). The nonlinear process can be blue-shifted through controlled cladding deposition (e.g. SiO$_2$) on the hybrid waveguide. This pulls the optical modes upwards, effectively mimicking an increase in LN film thickness. Conversely, red-shifting is achieved by thinning the printed PPLN, which pushes the modes downwards. To accomplish this, we use a modified RCA-1 clean (NH$_4$OH:H$_2$O$_2$ 1:4), a method known to etch lithium niobate while maintaining a good surface roughness\cite{zhuang2023high, yang2022monolithic} (measured RMS roughness = 0.33~nm, supplementary Fig.\ref{suppfig_RCAroughness}). Since only the +x crystal plane is exposed in our waveguide geometry, poled and unpoled domains etch uniformly, unaffected by etch anisotropy. Additionally, we measure a high etching selectivity of 30 with silicon oxide (compared to a typical selectivity of 1 for Ar-based physical etches), allowing for precise thinning of the nonlinear component without affecting the silicon photonics circuit. These post-printing modifications enable phase-matching shifts of over 100~nm in both directions (Fig.\ref{fig_PMtuning}a), fully correcting for fabrication-induced offsets and making deterministic phase-matching possible.

We validate this tuning concept by characterizing the phase-matching curves for different cladding thicknesses and etch depths (see Fig.\ref{fig_PMtuning}c). Starting from an operating wavelength of 1560~nm, we test oxide thicknesses on the order of a few tens of nanometers, achieving phase-matching across the entire C-band, down to 1520~nm. Conversely, thinning the LN by just a few nanometers significantly shifts the phase-matching to higher wavelengths, as demonstrated up to 1590~nm. In both cases, the circuit and monitor experience identical tuning, preserving their overlap and therefore ensuring reliable monitoring of a closed nonlinear circuit.

Based on these results, we propose a strategy for a wafer-scale quadratic nonlinear platform (Fig.\ref{fig_PMtuning}e). Through micro-transfer printing, a full silicon photonics wafer can be populated with PPLN films sourced from the same region of the LNOI wafer. Combined with the mature CMOS fabrication tolerances, this approach minimizes device non-uniformity and leads to nearly-identical phase-matching with a consistent offset from the target wavelength across the full wafer. Optical characterization of the nonlinear circuits through a monitor determines the remaining shift. Coarse tuning by cladding control (blue-shifting) or LN thinning (red-shifting) across the full wafer simultaneously aligns the phase-matching to the desired working point. By staying slightly blue-shifted, local heaters can individually fine-tune the remaining measured few-nanometer variations between devices. Ultimately, this approach yields an entire silicon photonics wafer with uniformly and deterministically phase-matched quadratic nonlinearities at the designed operating wavelength.


\subsection*{Printable entangled photon-pair sources}

Parametric optical nonlinearities are widely employed for quantum state generation, including squeezed states\cite{zhao2020near} and heralded single-photon sources\cite{wang2024progress}. However, the technological challenges involved with reproducible phase-matching of quadratic nonlinearities have hindered their use in large-scale integrated photonic circuits, with demonstrations limited to few interfering devices\cite{chapman2024chip}. Large-system implementations typically fall back on four-wave mixing (FWM) approaches with relaxed phase-matching requirements, often benefiting from the scalability of the silicon photonics platform \cite{aghaee2025scaling,psiquantum2025manufacturable}. Even so, second-order nonlinearities provide distinct advantages over FWM. Their higher efficiency greatly reduces the required pump power, thereby mitigating parasitic nonlinear effects such as Raman noise. Furthermore, the large spectral separation between the pump and generated photon state significantly simplifies filter design for broadband and high-extinction pump suppression.

Our nonlinear platform represents a major step towards integrating quadratic nonlinearities into large-scale photonic quantum systems. For instance, arrays of efficient photon-pair sources with identical phase-matching curves can be printed in front of large reconfigurable interferometers to generate complex entangled states (Fig.\ref{fig_SPDC}a). For that purpose, we characterize spontaneous parametric down-conversion in our nonlinear waveguides and establish printable photon-pair sources.

A Hanbury Brown-Twiss setup is used to measure the second-order correlation function g$^{(2)}$, as shown in Fig.\ref{fig_SPDC}b. A tunable pulsed Ti:Sapph laser at the second harmonic wavelength with a repetition rate of 80~MHz serves as pump and is coupled to the nonlinear waveguide through a lensed fiber. The generated photon pairs are coupled out and sent through a long-pass filter for pump suppression. The photons are then split with a 50:50 beam splitter and sent to superconducting nanowire single-photon detectors (SNSPD), which are connected to a time controller.

We measure the correlation function and investigate the coincidence-to-accidental ratio (CAR) and on-chip pair coincidence rate (PCR) as function of the pump power. The correlation histogram corresponding to a CAR of 1154 and an on-chip PCR of 28~kHz (with an on-chip average pump power of -40.0~dBm) is shown in Fig.\ref{fig_SPDC}c. Clear bunching is observed with repeating peaks at 12.5~ns spacing, matching the pump repetition rate. Adjusting the pump power through a variable attenuator reveals a linear dependence of the PCR as function of pump power (Fig.\ref{fig_SPDC}d), with CAR$\sim$PCR$^{-1}$ (Fig.\ref{fig_SPDC}e), indicative to the SPDC process. These results confirm efficient photon-pair generation, establishing printable quadratic photon-pair sources on the scalable silicon photonics platform.


\section*{Discussion}
As previously noted, micro-transfer printing allows for efficient material use due to the high density of devices on the LNOI source wafer. This leads to reduced thickness variations between devices and therefore improved reproducibility, where remaining shifts can be compensated through local heaters. We can verify the viability of this approach by estimating the temperature needed for full compensation.

For chip sizes of 1~cm~$\times$~1~cm, a 200~mm silicon photonics wafer can host approximately 270 chips. Keeping in mind that the PPLN devices would typically be only a small part of a larger photonic circuit, 10 devices per chip (i.e. per cm$^2$) would be a realistic estimate. Our source design can further be optimized to a density of 7 devices/mm$^2$, leading to an area of 4~cm$^2$ (equivalent to 2~cm~$\times$~2~cm) on the LNOI source to populate an entire 200~mm silicon photonics wafer. Said differently, a single 4~inch LNOI wafer is able to populate over ten 200~mm wafers. The phase-matching deviations over an entire silicon photonic wafer will then be dictated by the non-uniformity of a 2~cm~$\times$~2~cm LNOI region rather than the full wafer. A variation of 5~nm over the local region could then be completely compensated through a maximum temperature difference of 75~K (based on the sensitivity of Fig.\ref{fig_foundrySiPh}a and tuning rate of Fig.\ref{fig_PMtuning}d), which is well within the capacity of standard on-chip heaters\cite{parra2024silicon, yong2022power}. Improved LNOI manufacturing or pre-selection of uniform areas can reduce the required temperature tuning even further.

Micro-transfer printing offers several key advantages. As a back end of line (BEOL) integration technique, it leaves the CMOS processing of the silicon photonics wafer unaffected. The cladding can be locally etched to access the nitride waveguide, onto which the PPLN is printed. Therefore, the existing process design kits can be leveraged and coarse tuning of the phase-matching does not alter the silicon photonics circuit. Moreover, the integration can be scaled to larger volumes by accommodating an array of posts on the PDMS stamp and transferring many devices in parallel\cite{justice2012wafer}. Additionally, smart design of the exposed waveguide allows phase-matching across a range of wavelengths through a single generic PPLN design (see Fig.\ref{suppfig_genericPPLN}). This approach requires only one LNOI lithography mask set to support different circuit designs, making it both cost-effective and scalable. Furthermore, the current PPLN design can be improved by adding broadband mode couplers to reduce insertion losses, as has been demonstrated by previous works\cite{churaev2023heterogeneously, ruan2023high}.

In this work, we have demonstrated precise control and reproducibility of phase-matching, focusing on the SHG and SPDC processes. This approach is also applicable to other second-order processes, going from sum frequency generation to optical squeezing, and more complex circuits can be used, e.g. by incorporating PPLN into a resonator to achieve printable optical parametric oscillators. The micro-transfer printing method further enables co-integration with other components, such as mode-locked lasers\cite{cuyvers2021low}, high-speed modulators\cite{vanackere2023heterogeneous} and ultra-fast photodiodes\cite{maes2023high}, to construct complete single-chip photonic systems.


\section*{Conclusion}

Periodically poled LN rib waveguides offer highly efficient optical frequency conversion but suffer from extreme sensitivity to fabrication variations, limiting their scalability to larger systems. We present a scalable quadratic nonlinear platform that harnesses the maturity of existing CMOS processing by heterogeneously integrating periodically poled LN onto silicon photonics. A generic PPLN design enables phase-matching on two distinct platforms with nonlinear efficiencies comparable to LNOI rib waveguides. A dual-waveguide approach is introduced to allow monitoring in closed optical nonlinear circuits and reproducibility is shown over a complete 200~mm silicon photonics wafer radius. A tuning mechanism based on controlled waveguide modifications over the full wafer simultaneously leads to blue- and red-shifting, fully compensating any fabrication-induced offsets. Based on these results, a wafer-scale strategy is formulated to obtain reproducible and deterministic phase-matching over the full wafer. Given its potential in large-scale quantum systems, the SPDC process is characterized in the proposed platform, establishing printable entangled photon-pair sources. 

This work marks an important step towards scalable quadratic nonlinearities, paving the way for their integration into large-scale optical nonlinear systems and facilitating high-volume manufacturing of highly-efficient frequency converters.


\bibliography{PPLN_references}
\bigskip


\section*{Methods}
\noindent\textbf{Fabrication of the nonlinear waveguide}

\noindent The fabrication starts from an X-cut LNOI wafer (300~nm LN/ 2~$\upmu$m SiO$_2$ / LN substrate, NanoLN). The first step is periodic poling of the LN, for which electrodes (50~nm Cr/ 100~nm Au) are deposited through a sequence of electron-beam lithography, electron-beam evaporation and a liftoff process. The electrodes are covered with an insulating oil and the LN is poled through a single custom electric waveform, reaching a maximum voltage of 450~V. The quality of the inverted ferro-electric domains is characterized through PFM. After poling, the electrodes are removed through selective wet chemical etching. 

Next, an amorphous silicon hard mask is deposited, which is patterned into rectangular films (40~$\upmu$m~$\times$~1~mm) through a combination of UV-photolithography and reactive ion etching (RIE). The LN layer is then fully etched using an Ar-based RIE process, exposing the buried oxide layer underneath. The hard mask is subsequently selectively removed with a KOH solution. The PPLN films are reinforced through a patterned photoresist encapsulation, after which a long HF-based etch removes the oxide layer underneath the full PPLN film. This leaves the devices suspended with only a few LN attachment points to the side that keep them connected to the substrate.

In parallel, a silicon photonics chip is prepared. For the in-house platform, a uniform SiN chip (300~nm SiN/ 3.3~$\upmu$m SiO$_2$/ Si substrate) is patterned through electron-beam lithography and RIE. For the foundry platform, a 200~mm silicon photonics wafer is provided by Imec, containing 300~nm thick SiN waveguides that are planarized through SiO$_2$ cladding followed by chemical-mechanical polishing. Before transfer printing, the chips are covered with photosensitive benzocyclobutene (BCB) where UV-lithography is used to remove the BCB directly above the nitride waveguides. This allows for adhesive bonding of the PPLN film on the sides, while having direct bonding on top of the waveguide. As a result, BCB thickness variations will not influence the phase-matching as the optical mode does not feel the BCB.

The PPLN devices are picked and printed onto the prepared silicon photonics chips with a polydimethylsiloxane (PDMS) stamp using a commercial micro-transfer printing tool (X-Celeprint $\upmu$TP-100). The photoresist encapsulation is removed with acetone and a brief O$_2$ plasma. Lastly, the BCB is cured at 280$\degree$C for one hour in vacuum, leaving the devices ready for measurement.\\

\noindent\textbf{Phase-matching characterization}
To characterize the phase-matching, the silicon photonics chips are cleaved to define facets of the silicon nitride waveguides. Coupling to the waveguides is achieved through two types of lensed fibers, which are single-mode at 1550~nm and 780~nm, respectively. The silicon photonics chips are designed symmetrically such that the fiber-to-PPLN losses can be characterized at both fundamental and second harmonic wavelength. Knowing the coupling losses at both wavelengths, the pump and generated second harmonic power are measured with calibrated detectors, allowing the corresponding powers in the hybrid waveguide to be calculated. The normalized nonlinear efficiency then results from:
\begin{minipage}{\linewidth}
    \vspace{0 pt}
    \begin{equation*}
        \eta = \frac{P_{SH}}{P_F^2 \cdot L^2}
        \vspace{0 cm}
    \end{equation*}
    \vspace{0 pt}
\end{minipage}
where $P_{F}$ and $P_{SH}$ are the fundamental and second harmonic power in the PPLN/SiN waveguide, respectively, and L is the nonlinear interaction length. Although the PPLN film has a total length of 1~mm, the interaction length is limited to 0.9~mm as 50~$\upmu$m at each end remains unpoled where SiN waveguide transitions are located.


\section*{Data Availability}
\noindent Data used in this study are available from the corresponding author
upon request.

\section*{Acknowledgments}

The authors would like to acknowledge the contribution of Imec’s 200~mm pilot line for development and fabrication of the silicon photonics wafer. We want to thank and acknowledge funding by the Research Foundation Flanders (FWO) (11H6723N, 1S69123N), FWO-SBO through the BeQuNet grant (S008323N), Fonds de la Recherche Scientifique (MIS F.4506.20) and the FWO and F.R.S.-FNRS under the Excellence of Science (EOS) program (40007560). St\'ephane Clemmen is a research associate of the Fonds de la Recherche Scientifique (FNRS)

\section*{Author contributions}

T.V., S.C. and B.K. conceived the idea for the project. T.V., J.D.W., L.D.J. have developed the processes (poling and transfer printing), and fabricated the nonlinear waveguides. T.V., E.V. and L.D.J. designed and numerically simulated the components. T.V., L.D.J. and E.V optically characterized the waveguides. T.V. and J.D.W. measured and characterized the SPDC process. S.J., P.V., N.S., D.B., M.D., F.F., P.A., S.B. and J.V.C. have developed Imec's 200~mm silicon photonics platform and provided the wafer. T.V. prepared figures, data and wrote the manuscript with input from co-authors. D.V.T., G.R., S.C. and B.K. supervised the project.

\section*{Competing interests}
\noindent The authors declare no competing interests.\\


\clearpage

\newgeometry{a4paper, left=1in, right=1in, top=1in, bottom=1in}  
\fontsize{10pt}{20pt}\selectfont  
\onecolumngrid 

\renewcommand{\thesection}{\arabic{section}} 
\renewcommand{\thepage}{S\arabic{page}} 

\setcounter{section}{0} 
\setcounter{page}{1} 
\setcounter{figure}{0}

\renewcommand{\figurename}{FIG.}
\renewcommand{\thefigure}{S\arabic{figure}} 

\begin{center}
    {\LARGE \textbf{Supplementary Material}}\\[1cm]
\end{center}

\section{Insertion loss of the nonlinear component}

\begin{figure}[h]
\centering
\includegraphics[width=\linewidth]{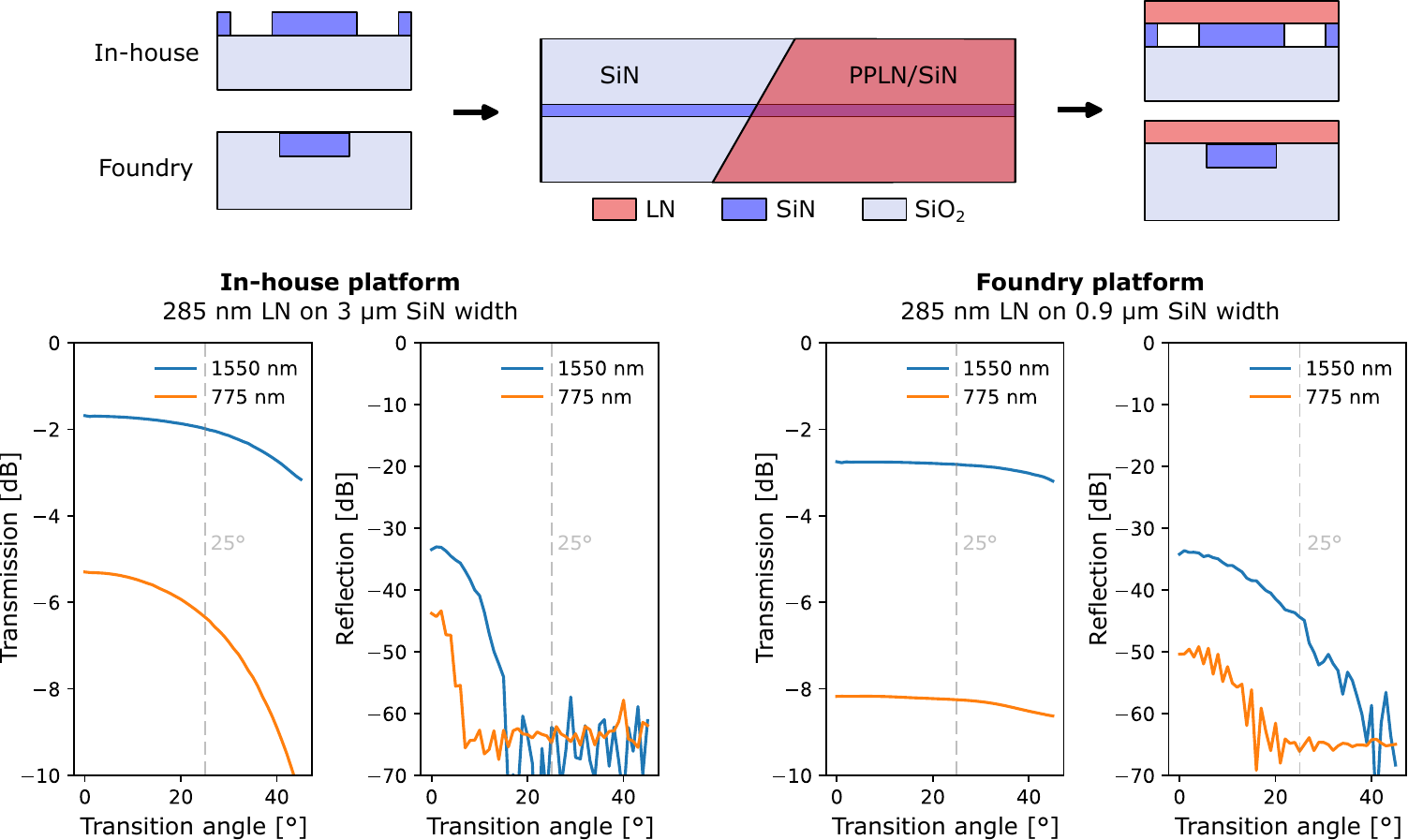}
\caption{\textbf{Losses at the angled transitions.} The LN film is introduced at an angle when transitioning from SiN to the hybrid PPLN/SiN waveguide. Simulations on the in-house (left) and foundry (right) platform are shown, with the corresponding transmission and reflection as function of the angle. }
\label{suppfig_angledtransition}
\end{figure}

The nonlinear component features an angled transition at 25$\degree$ to minimize unwanted Fabry-Perot fringes in the spectral response. Simulations on both silicon photonics platforms are shown in Fig.\ref{suppfig_angledtransition}. 

On the in-house platform, the SiN waveguide width is 3~$\upmu$m at the transition and tapers down to the phase-matching width beneath the film (as seen in Fig.\ref{fig_process flow}d). The LN film thickness used in the simulations is 285~nm, which is closer to the fabricated devices compared to the desired 300~nm. This discrepancy originates from the LNOI manufacturing. Simulations indicate that the reflections can be reduced significantly ($\sim$~30~dB) through an angled transition, while only slightly affecting the transmission. Based on these results, we selected an angle of 25~$\degree$. 

On the foundry platform, the SiN waveguide is 900~nm wide over the full device, corresponding to the phase-matching width. With future work in mind, the exposed nitride waveguides were designed to incorporate a 3~mm PPLN film, in contrast to the current 1~mm.
Therefore, having wider SiN widths at both transitions is not possible. The choice was made to print the 1~mm such that both transitions of the hybrid waveguide were identical, which is important to estimate the nonlinear efficiency. As seen in the simulations, this leads to a small increase in losses compared to the in-house platform. By implementing mode couplers in future designs, these transition losses can be mitigated, without affecting the developed phase-matching functionalities.

Besides the transition losses, propagation losses in the hybrid waveguide also contribute to the overall loss of the component. However, as only a single interaction length was measured in this work, these cannot be separated from the transition losses. Given the limited interaction length of 1~mm, the losses over the full component are dominated by angled transition losses, corresponding to above simulations.

\begin{figure}[h]
\centering
\includegraphics[width=\linewidth]{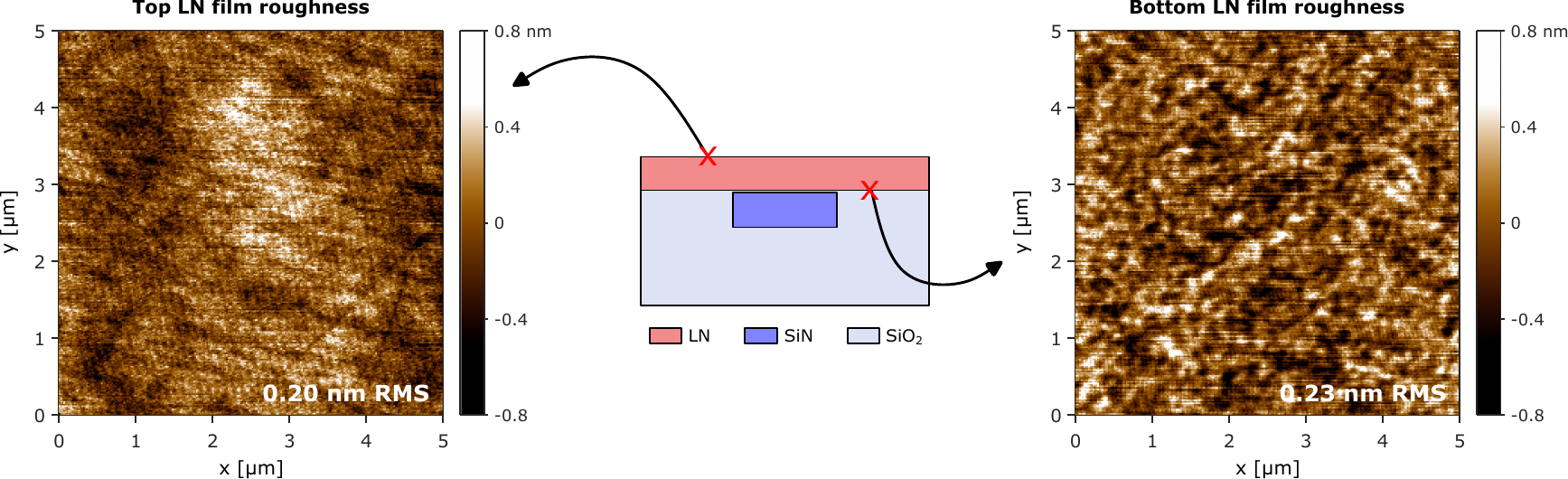}
\caption{\textbf{Surface roughness of transfer printed LN film.} AFM measurements of top and bottom (by picking and flipping the LN film with a PDMS stamp) of the LN films. Good RMS roughnesses of 0.20~nm and 0.22~nm are measured, respectively, similar to the original LNOI wafer.}
\label{suppfig_LNroughness}
\end{figure}

To assess the potential impact of the transfer printing process on propagation losses, we measured the surface roughness of the LN film (Fig.\ref{suppfig_LNroughness}).  

The micro-transfer printing process has been optimized to maintain the original surface roughness of the LNOI wafer. AFM measurements reveal an RMS roughness of 0.20~nm on top and 0.22~nm on the bottom of the film (measured by picking and flipping the film with a PDMS stamp). These results show that the original roughness is maintained, suggesting that low propagation losses should be feasible. 

Thus, the primary source of insertion loss in the component stems from the angled transitions, which can be mitigated by incorporating mode couplers in future designs. The minimal surface roughness of the printed LN film suggests that low propagation losses should be achievable, further improving overall device performance.

\section{Observation of higher-order harmonics}

\begin{figure}[h]
\centering
\includegraphics[width=\linewidth]{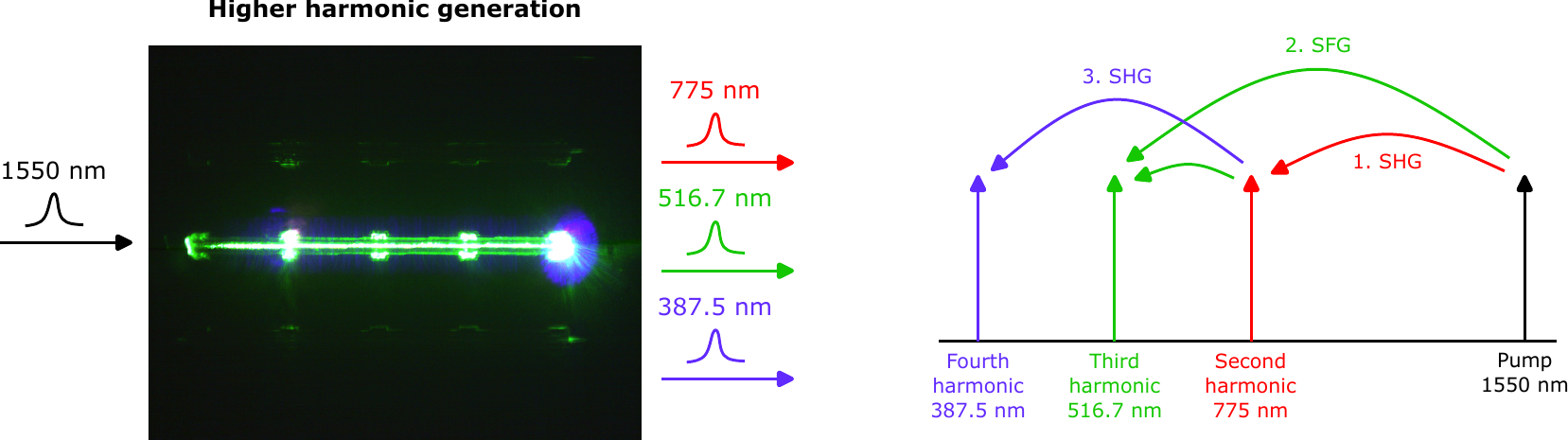}
\caption{\textbf{Imaging of higher-order harmonics.} Using a camera with a standard NIR-blocking filter, green and blue light generation is imaged, localized in the hybrid waveguide.}
\label{suppfig_higherharmonics}
\end{figure}

It is well known that increasing the power in a phase-matched $\chi^{(2)}$ process can lead to cascaded nonlinear processes. We report the observation of higher-order harmonics localized within the hybrid waveguide (Fig.\ref{suppfig_higherharmonics}).

A PPLN/SiN waveguide is pumped with a C-band pulsed femtosecond laser at the phase-matching wavelength to reach high peak powers. The nonlinear waveguide is imaged through a CCD camera equipped with a NIR-blocking filter, which removes the second harmonic signal. A bright green and weaker blue light are observed inside the hybrid waveguide. Their absence in the silicon nitride before the component, in combination with the green buildup at waveguide input, suggests that the light originates from the PPLN component.

The green light is attributed to a sum-frequency process between the pump and the second harmonic, leading to a third harmonic at 516.7~nm. The blue light is attributed to a second harmonic generation process of the second harmonic, leading to the fourth harmonic at 387.5~nm.

Due to significant dispersion of the femtosecond pulse in the fibers used for edge coupling and the challenge of measuring coupling losses at the generated wavelengths, a quantitative estimation of the conversion efficiencies is challenging. Nevertheless, the localization within the 1~mm PPLN waveguide points to a strong nonlinear response, related to the $\chi^{(2)}$ component. This further confirms the integration of the second-order nonlinearity on a silicon photonics platform.

\begin{figure}[h]
\centering
\includegraphics[width=\linewidth]{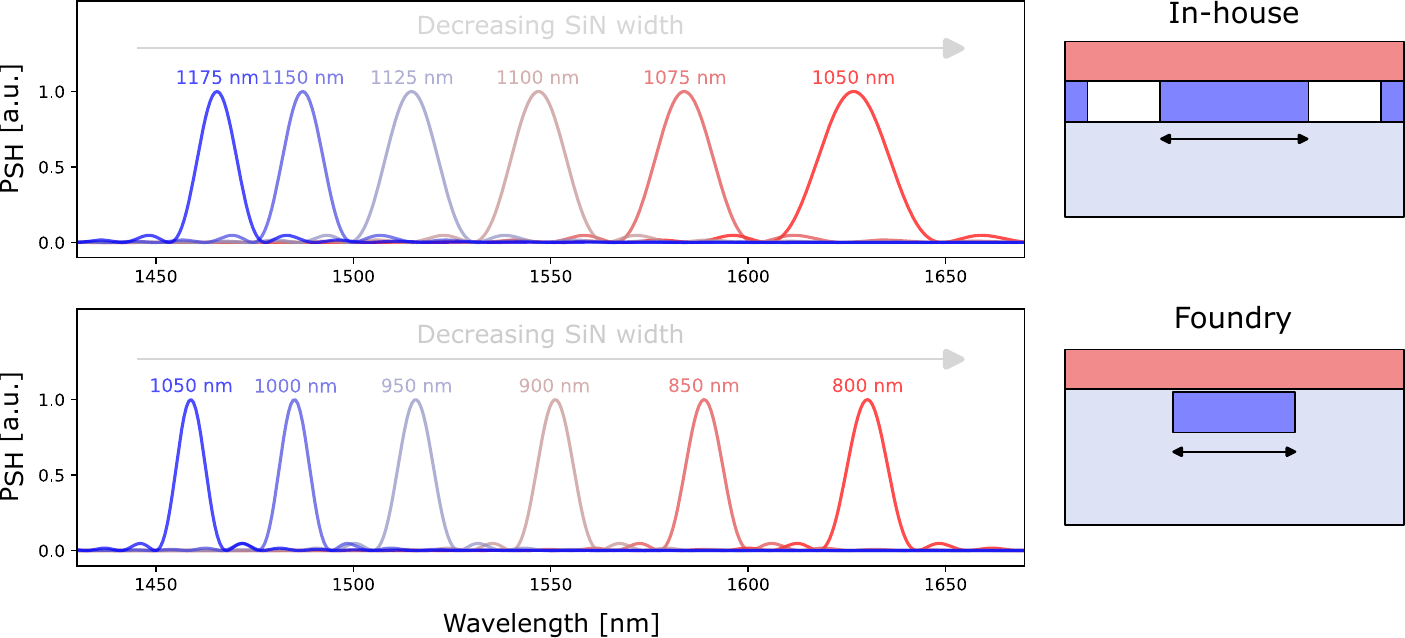}
\caption{\textbf{Phase-matching accross distinct platforms at multiple wavelengths with a generic PPLN design} Simulations of obtained phase-matching for in-house and foundry platform for several wavelengths by changing the nitride width, while keeping the PPLN design unchanged.}
\label{suppfig_genericPPLN}
\end{figure}

\section{Generic PPLN design for cross-platform and multi-wavelength phase-matching}

The micro-transfer printing method enables the use of a single generic PPLN film design to achieve phase-matching across different silicon photonics platforms, as demonstrated in Fig.\ref{fig_SHGcharacterisation}a and \ref{fig_SHGcharacterisation}b. We support this claim by simulating the phase-matching curves for a range of geometries that share the PPLN design (i.e. poling period and LN film thickness), but vary in SiN width (see Fig.\ref{suppfig_genericPPLN}).

The simulations assume a constant LN thickness of 290~nm and poling period of 3.5~$\upmu$m. By varying the silicon nitride width, the phase-matching is swept from 1450~nm to 1650~nm, with the corresponding SiN widths labeled above their respective curves. The results show phase-matching over 200~nm for both platforms, demonstrating the concept of a generic PPLN design. Due to this versatility, the LNOI source wafer can be filled with a single optimized design which can address phase-matching over different platforms with a range of possible wavelengths.

\section{Fabrication non-uniformity and non-ideal periodic poling}

\begin{figure}[h]
\centering
\includegraphics[width=\linewidth]{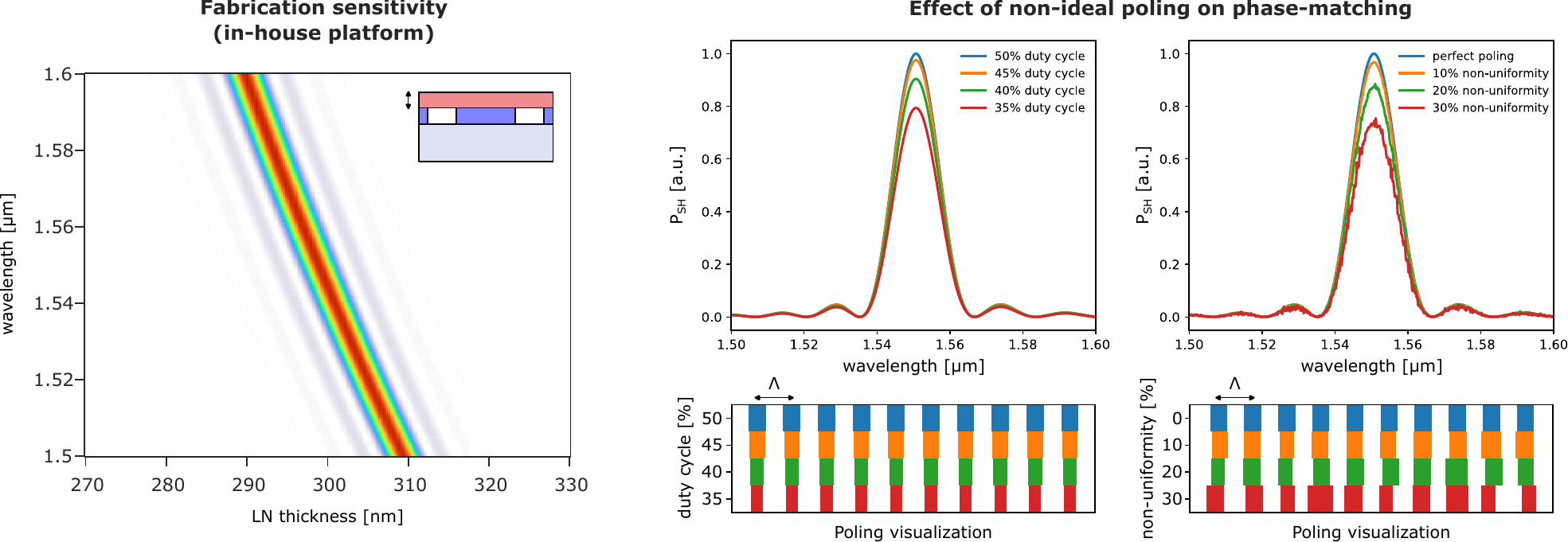}
\caption{\textbf{In-house fabrication sensitivity and effect of non-ideal periodic poling.} Left: Numerical simulations of the sensitivity to LN thickness variations of the in-house platform (similar to Fig.\ref{fig_foundrySiPh}a for foundry platform). Right: Effect of non-ideal poling for smaller duty cycles and non-uniform poling, with a visualization of the poled domains corresponding to the simulations.}
\label{suppfig_nonidealfabrication}
\end{figure}

Fabrication variations are known to be the primary cause for unexpected shifts and deviations from the sinc curve of the phase-matching. In our hybrid waveguide, the dominating variations arise from the original LN film thickness. The sensitivity to these variations is simulated and given in Fig.\ref{fig_foundrySiPh}a (foundry) and Fig.\ref{suppfig_nonidealfabrication} (in-house), resulting in 2.8 nm and 5 nm, respectively, for every nanometer difference in thickness. 
Besides variations in geometry, non-ideal poling will also reduce the efficiency of the nonlinear process. We investigate the effect of smaller duty cycles and non-uniform poled domains on the phase-matching curve, based on the model of Wu et al.$^1$. Fig.\ref{suppfig_nonidealfabrication} shows the results and a visualization of the inverted domains. A duty cycle down to 35$\%$ will reduce the nonlinear efficiency to 80$\%$ of the ideal scenario. Similar, a non-uniformity of 30$\%$ (where domain walls are randomly displaced by up to 30\% of half the poling period) keeps an efficiency of 70$\%$ relative to perfect poling. Random non-uniformity will also add noise-like variations to the curve, though the phase-matching wavelength remains unchanged. This implies that non-uniformity will be the primary cause of spectral shifts in phase-matching, as expected. Furthermore, the monitor waveguide introduced in Fig.\ref{fig_foundrySiPh}c will still have an identical phase-matching wavelength as the circuit waveguide for non-ideal poling, albeit at slightly different absolute nonlinear efficiency.\\

{\small $^1$ Wu, X. \textit{et al.} Broadband second-harmonic generation in step-chirped periodically poled lithium niobate waveguides. \textit{Optics Letters} \textbf{47}, 1574–1577 (2022).}\\

\section{Modified RCA-1 etch: surface roughness}

\begin{figure}[h]
\centering
\includegraphics[width=0.4\linewidth]{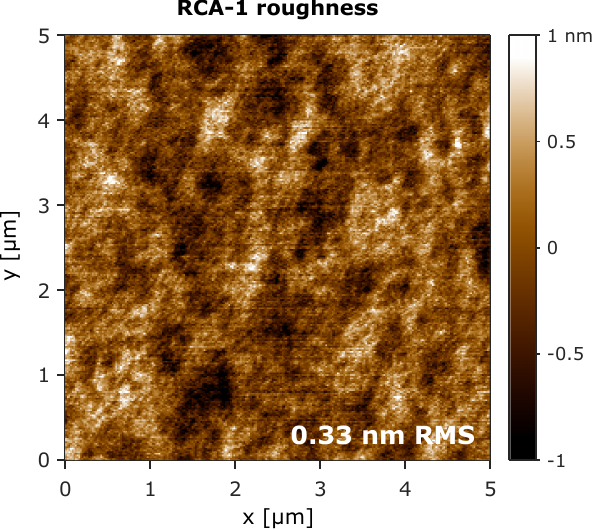}
\caption{\textbf{Roughness measurement of the modified RCA-1 etch.} AFM measurement of 250~nm etched X-cut LN, giving an RMS roughness of 0.33~nm. }
\label{suppfig_RCAroughness}
\end{figure}

A modified RCA-1 etch (NH$_4$OH:H$_2$O$_2$ 1:4) is used to thin the LN film after printing, red-shifting the phase-matching wavelength. The etch rate can be controlled by adjusting the temperature: at 85$\degree$C, 300~nm is etched in 30~minutes, while at room temperature, the rate slows down to 3~nm in 30~minutes. The latter is desired for precise control of the thickness for tuning, as demonstrated in Fig.\ref{fig_PMtuning}c. To assess the roughness, an AFM measurement was performed on an X-cut LNOI chip that started at 300~nm and was etched down to 50~nm, with the result shown in Fig.\ref{suppfig_RCAroughness}. An RMS roughness of 0.33~nm was measured, demonstrating low-roughness thinning of the lithium niobate.

\end{document}